# Plasmonic Nanostars with Hot Spots for Efficient Generation of Hot Electrons under Solar Illumination

*Xiang-Tian Kong,\* Zhiming Wang,\* Alexander O. Govorov\**

Dr. X.-T. Kong
Institute of Fundamental and Frontier Sciences and State Key Laboratory of Electronic Thin Films and Integrated Devices, University of Electronic Science and Technology of China, Chengdu 610054, China
Department of Physics and Astronomy, Ohio University, Athens, Ohio 45701, United States
E-mail: kt1729@gmail.com
Prof. Z. Wang
Institute of Fundamental and Frontier Sciences and State Key Laboratory of Electronic Thin Films and Integrated Devices, University of Electronic Science and Technology of China, Chengdu 610054, China
E-mail: zhmwang@gmail.com
Prof. A. O. Govorov
Department of Physics and Astronomy, Ohio University, Athens, Ohio 45701, United States
E-mail: govorov@helios.phy.ohiou.edu

Nanostars (NSTs) are spiky nanocrystals with plasmonic hot spots. In this study, we show that strong electromagnetic fields localized in the nanostar tips are able to generate large numbers of energetic (hot) electrons, which can be used for photochemistry. To compute plasmonic nanocrystals with complex shapes, we develop a quantum approach based on the effect of surface generation of hot electrons. We then apply this approach to nanostars, nanorods and nanospheres. We found that that the plasmonic nanostars with multiple hot spots have the best characteristics for optical generation of hot electrons compared to the cases of nanorods and nanospheres. Generation of hot electrons is a quantum effect and appears due to the optical transitions near the surfaces of nanocrystals. The quantum properties of nanocrystals are strongly size- and material-dependent. In particular, the silver nanocrystals significantly overcome the case of gold for the quantum rates of hot-electron generation. Another important factor is the size of a nanocrystal. Small nanocrystals are more efficient for the hot-electron generation since they exhibit stronger quantum surface effects. The results of this study can useful for designing novel material systems for solar photocatalytic applications.

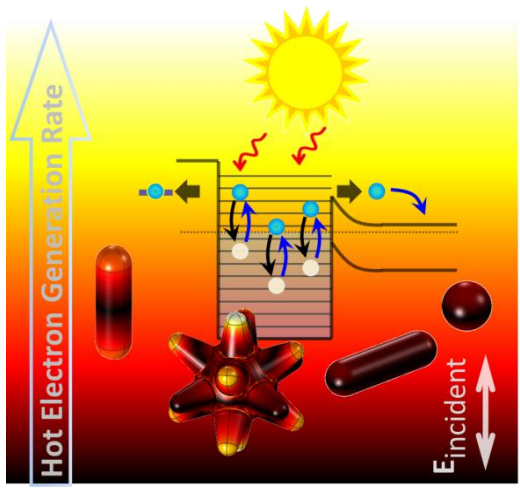

1. Introduction

Hot plasmonic electrons attract currently lots of interest. Such electrons are generated by incident radiation in plasmonic nanostructures and can be used for photochemistry[1,2,3,4,5,6,7,8,9,10] and optoelectronics.[11,12,13,14,15,16,17,18,19,20, 21,22] The challenges to create useful devices based on the plasmonic photoelectric effect lie in the short relaxation times of excited electrons in metals and in the small numbers of energetic carriers generated in typical nanocrystals. [23] One possible way to overcome the above challenges is to use special designs of nanocrystals and, in particular, to utilize the nanostructures with hot spots. [24,25,26,27] Hot spots in plasmonic nanostructures are small spaces where an incident electromagnetic field becomes strongly amplified due to the resonant plasmonic effect. It was also observed [24,26,27] and calculated [25] that the plasmonic hot spots in metal nanostructures can generate large numbers of energetic electrons. The design of a plasmonic nanocrystal is crucial to create plasmonic hot spots that generate efficiently energetic electrons. In this study, we will address this problem of hot spots and energetic electrons.

Here we investigate theoretically the process of generation of hot electrons in nanocrystals with various shapes, such as nanostars (NSTs), nanorods (NRs) and spherical nanoparticles (NPs). The focus will be on the role of plasmonic hot spots in such nanocrystals (NCs). In this study, we show that plasmonic NSTs with multiple hot spots are very efficient for generation of energetic electrons. We also compare NSTs with NRs and NPs and discuss two material systems, gold and silver. Silver NCs exhibit much larger rates of generation since the plasmon enhancement in such silver NCs is much stronger than that in the case of gold. The physical reason for the high rates of generation of hot carriers in the silver NCs are the following: (1) The long mean free path of electrons and (2) the narrow and intensive plasmonic resonances. The effect of generation of hot electrons is an essentially quantum effect and comes from the optical absorption processes near the surfaces of a NC. Such surface absorption is, of course, efficient only in relatively small NCs. We, therefore, examine the quantum efficiencies and the quantum plasmonic parameter for NCs of various sizes and shapes. Using such calculations, we show the quantum nature of the generation of energetic carriers in NCs in general.

We also should note that the formalism of the paper focuses on the quantum intraband transitions in the NCs made of gold and silver. Such intraband transitions dominate the optical responses of the plasmonic NCs in the spectral intervals $\lambda > 500 nm$ (gold) and $\lambda > 400 nm$ (silver). Simultaneously, these spectral intervals are most interesting for the plasmonic effects since the plasmonic peaks in these spectral regions appear to be narrow and strong. [23]

Regarding the role of the interband transitions and generation of hot holes in the d-band, one can look into recent review papers (e.g. Ref. [23]).

Theoretically, the problem of hot electrons has been addressed using several methods: Density matrix formalism combined with time-dependent DFT, perturbative approach for the injection currents, Fermi's golden rule, non-equilibrium Green functions, quantum kinetic approaches, etc.[14,25,26,28,29,30,31,32,33,34] Since this paper deals with NCs of complex shapes, we will develop and employ a simplified quantum method based on the surface density of generation of hot carriers in the plasmonic NCs. In particular, our formalism will involve integration of the rate of generation over the surface area of a complex NC with hot spots. In contrast to the field theories, which quantize plasmonic excitations, [35,36] we look at the electronic structure of the plasmon and calculate the number of high-energy electrons in the excited plasmonic wave.

## 2. Electromagnetic Models of Nanocrystals and Quantum Formalism for the Generation of Hot Electrons

**Figure 1** shows the electromagnetic models of the NCs computed in this study. In particular, we will look at spherical nanoparticles (NPs), nanorods (NRs) and nanostars (NSTs). The NSTs are especially interesting for the plasmonic effects since they have multiple hot spots at the tips.[10,27,37,38,39,40] It was also shown experimentally and theoretically in Refs.[10,27,37,38,39,40] that the geometrical parameters of the tips in the NSTs strongly influence the positions of plasmonic resonances is such NCs. In our study, we first calculate classical optical properties of nanocrystals using the COMSOL software and the empirical dielectric constants of metals (gold and silver).[41] For the dielectric constant of the matrix, we take $\varepsilon_0 = 2$; this number corresponds to a water matrix. In particular, we are interested in the absorption cross section and in the strength of the plasmonic surface fields. For the power of absorption and for the related absorption cross section, we have

$$Q_{abs} = \left\langle \int_{NPs} dV \, \mathbf{j} \cdot \mathbf{E} \right\rangle_{time} = \text{Im}(\varepsilon_{metal}) \frac{\omega}{2\pi} \int_{NPs} dV \, \mathbf{E}_\omega \cdot \mathbf{E}_\omega^*,$$
$$\sigma_{abs}(\omega) = \frac{Q_{abs}}{I_0},$$
(1)

where $\mathbf{j}_\omega$ is the dynamic current in the NCs, $\mathbf{E}_\omega$ is the complex amplitude of the electric field inside the NC, and $\varepsilon_{metal}$ is the dielectric constant of the metal. The parameter $I_0$ is the incident flux of light energy given by the following equation:

$$I_0 = \frac{c_0 \sqrt{\varepsilon_0}}{2\pi} \cdot E_0^2$$

where $E_0$ is the amplitude of the electric field in the incident electromagnetic wave. Since we work with complex numbers, the physical value of the electric field is the real part of the complex field:

$$\mathbf{E} = \mathbf{E}_\omega \cdot e^{-i\omega t} + \mathbf{E}_\omega^* \cdot e^{+i\omega t}$$

The physical parameter responsible for the effect of generation of hot plasmonic electrons is the normal-to-surface component of the electric field near the surface of a nanocrystal (Figure 1). Then, we can introduce the related integral parameter, which also describes the appearance of the hot spots in a nanocrystal,

$$Enh_S = \frac{1}{S_{NC}} \int_{S_{NC}} \frac{|E_{normal}|^2}{E_0^2} ds$$
(2)

where the integral is taken over the NC surface; the function is the normal field at the metal-matrix interface inside the NC and the parameter is the surface area of the nanocrystal. In fact, the parameter (2) can be regarded as an enhancement factor for the normal electric fields at the surface inside a NC.

The mechanism of generation of hot electrons, which we involve in this study, comes from the quantum optically-induced transitions of electrons near the surface of the NCs.[23,25,28] Near the surface of a NC, a linear momentum of electron is not conserved and the electromagnetic field is able to generate excited electronic states with large energies. Such energetic (hot) electrons generated near the surface represent a surface effect. Simultaneously, the plasmonic field creates a large number of electrons with small excitation energies in the bulk of a NC. The classical Drude model describes such bulk excitations. These bulk excitations also form the dissipative currents creating Joule heat inside a NC. **Figures 1** and **2** illustrate the phenomena of hot and Drude electrons. Hot electrons are generated near the surfaces (Figure 1b) and their energies extend over the whole interval $E_F < \varepsilon < E_F + \hbar\omega$, which is allowed by the conservation-of-energy law (Figure 2b). Here we denoted the Fermi energy as $E_F$. The Drude electrons induce the so-called frictional heating of a NC and have small excitation energies, $\sim k_B T$. The nonequilibrium state of a NC can be described by the distribution function $\delta f(\varepsilon) = dN_{excited}/d\varepsilon$, where $N_{excited}$ is the number of excited electrons in a NC and $\varepsilon$ is the energy of electron. As an example, Figure 2b shows the typical distribution function of excited electrons in an optically-driven NC. The positive values of $\delta f(\varepsilon)$ in Figure 2b correspond to excited electrons above the Fermi level, whereas the negative values of $\delta n(\varepsilon)$ describe the creation of holes in the Fermi sea. To compute the distribution in Figure 2b, we used the quantum formalism from Refs. [25,28,42], which was applied to a spherical NP in Ref. [43]. In this figure, we also show the over-barrier electrons as a red region. These energetic electrons can be injected from the NC.

In the past, we have applied the fully quantum formalism of hot electrons to slabs, spheres and cubes. [25,28,42] However, it is challenging to perform such fully quantum calculations for NCs with arbitrary complex shapes because of the complexity of single-particle wave functions. Nevertheless, it may also be possible to find certain approximate methods for complex NCs. Here we like to show how to treat the nanocrystals with complex shapes, such as NSTs and NRs. First, we should define the surface density of hot electrons, $\sigma_{high-energy\,electrons}(\theta,\varphi)$; here $\theta$ and $\varphi$ are the spherical angles. For this, we now use the formula derived by us in Ref. [25]:

$$\frac{d^2 N_{high-energy}}{ds d\varepsilon} = \left|eE_{normal}\right|^2 \times \frac{2}{\pi^2} \times \frac{E_F^2}{\gamma_\varepsilon} \frac{1}{(\hbar\omega)^4} \quad (3)$$

where $d^2 N_{high-energy}/ds d\varepsilon$ is the number of high-energy electrons for unit area and energy; $\gamma_\varepsilon = \hbar/\tau_\varepsilon$ is the rate for energy relaxation of single electrons. The rates of generation of hot electrons, which will be calculated within our model, will not depend on the parameter $\tau_\varepsilon$. The typical energy relaxation times in gold NCs are in the range $\tau_\varepsilon = 0.1 - 0.5\,ps$. [44] The function (3) has the dimensionality of $1/erg\cdot cm^2$ and, therefore, the total rate of generation of hot electrons per unit surface area should be calculated as

$$\frac{dR_{high-energy}}{ds} = \frac{1}{\tau_\varepsilon}\frac{d^2 N_{high-energy}}{ds d\varepsilon}\cdot(\hbar\omega - \Delta E_b) \quad (4)$$

Here $\Delta E_b$ is the height of the electronic barrier between the plasmonic NC and the semiconductor contact (Figure 1b). The barrier energy $\Delta E_b$ can also be the energy spacing between the Fermi level and the level of adsorbed molecule as shown in Figure 1b. In our calculations, we take $\Delta E_b = 1\,eV$ that is a typical number for barrier heights in experimental systems. Equation (4) assumes that the distribution function $\delta f(\varepsilon)$ is flat in the region $E_F + \Delta E_b < \varepsilon < E_F + \hbar\omega$, which is a valid assumption (Figure 2b). Then, we calculate the total rate of generation of hot electrons in a NC as an integral:

$$Rate_{high-energy} = \int_{S_{NC}} \frac{dR_{high-energy}}{ds} ds \quad (5)$$

This integral should be taken over the NC surface. Using Equation (3)-(5), we arrive to the final equation for the rate:

$$Rate_{high-energy} = \frac{2}{\pi^2} \times \frac{e^2 E_F^2}{\hbar} \frac{(\hbar\omega - \Delta E_b)}{(\hbar\omega)^4} \int_{S_{NC}} |E_{normal}(\theta,\varphi)|^2 ds \quad (6)$$

This integral (6) has the dimensionality of $1/s$ and gives the total rate of optical generation of over-barrier electrons in a plasmonic NC. Equation (6) is the central formula of this study. Below, we will apply this formula to the three geometries, NPs, NRs and NSTs.

For the case of a spherical NP, we can easily derive an analytical equation for the rate from the general Equation (6). The total electric field inside a small plasmonic sphere is well known:

$$\mathbf{E}_{inside} = \mathbf{E}_0 \frac{3\varepsilon_0}{2\varepsilon_0 + \varepsilon_{metal}}$$

where $\mathbf{E}_0$ is the external electric field of light. Then, Equation (6) yields

$$Rate_{high-energy} = \frac{2}{\pi^2} \times \frac{e^2 E_F^2}{\hbar} \frac{(\hbar\omega - \Delta E_b)}{(\hbar\omega)^4} \frac{4\pi}{3} R_0^2 \left|\frac{3\varepsilon_0}{2\varepsilon_0 + \varepsilon_{metal}}\right| \cdot E_0^2 \quad (7)$$

Since the hot-electron generation effect is due to the surface of a NC, we should expect that $Rate_{high-energy} \sim S_{NC}$, where $S_{NC}$ is the NC surface area. Indeed, we see this dependence in Figure 2c and in Equation (7). We also need to compare our simplified formalism based on Equation(6) with the exact quantum formalism from Ref. [25] . Figure 2c shows this comparison. The dots in Figure 2c are the result of the full quantum calculation according to the formalism of Ref. [25] , whereas the curve comes from Equation (7). We see excellent agreement between the simplified formalism used in this study (Equation(6)) and the quantum theory of Ref. [25] . In the following sections, we will be applying the present formulism based on Equation (6) to the cases of NSTs and NRs.

It is interesting to define a quantum efficiency of hot electron generation. We can do it in the following way:

$$Eff_{high-energy} = \frac{Rate_{high-energy}}{Rate_{absorption, photons}} = \frac{Rate_{high-energy}}{Q_{abs,tot} / \hbar\omega} \qquad (8)$$

where $Rate_{absorption, photons}$ is the rate of absorption of photons by the NC and $Q_{abs,tot}$ is the total power of absorption. Of course, the photon rate and the absorbed power are related via the simple equation

$$Rate_{absorption, photons} = \frac{Q_{abs,tot}}{\hbar\omega}$$

The total rate of energy dissipation in a NC has two terms: $Q_{abs,tot} = Q_{abs} + Q_{abs,quantum}$, where $Q_{abs}$ is given by the classical formalism (Equation(1)) and $Q_{abs,quantum}$ is the quantum term coming from the generation of hot electrons at the surfaces of a NC. Within our approach, the quantum contribution should be calculated as

$$Q_{abs,quantum} = \frac{2}{\pi^2} \times \frac{e^2 E_F^2}{\hbar} \frac{1}{(\hbar\omega)^2} \int_{S_{NC}} |E_{normal}(\theta,\varphi)|^2 \times ds \qquad (9)$$

The derivation for this equation is given in Supporting Information. We may also introduce another interesting parameter that describes the role of the quantum effects in absorption of a NC:

$$QP = \frac{Q_{abs,quantum}}{Q_{abs,tot}} \qquad (10)$$

The above quantum parameter ($QP$) is a ratio between the quantum dissipation and the total dissipation in a NC. We also should note that these parameters ($Eff_{high-energy}$ and $QP$) are related and describe the role of quantum hot-electron generation of energetic electrons in the plasmonic kinetics of a NC.

The above formalism gives a convenient approach to compute the effect of generation of hot electrons in NCs with arbitrary shapes. This approach is a combination of classical electrodynamics and quantum mechanics. Classical electrodynamics is used to compute local electromagnetic fields, whereas quantum mechanics is applied via the equation for the local surface rate of generation of energetic carriers. The main limitation of our approach is in the sizes of NCs. Our approach is valid when all characteristic sizes of the NC are greater than the lattice period of the bulk crystal.

## 3. Results for Spherical Nanoparticles, Nanorods and Nanostars

We now turn to the numerical results for complex plasmonic NCs. The most interesting geometry is multi-spike plasmonic NSTs shown in insets of **Figure 3**. In this sequence of the NSTs, we vary the spike length, but we keep the volume constant. The effective diameter of all plasmonic objects in Figure 3 is equal to 30 nm. The absorption cross section dramatically develops as we move from the sphere (NST#9) to the NST with the longest spikes (NST#1).

The main dipolar plasmon peak exhibits the expected red shift with increasing the spike length. Simultaneously, the absorption cross section and the plasmonic enhancement factor (Figure 3a,b) strongly grow. The reason is the well-known behavior of the gold nanocrystals coming from the properties of bulk dielectric constant of gold. In bulk gold, any excitations in the wavelength interval <600 nm experience strong damping due to the inter-band transitions[45] and, therefore, all plasmonic effects become strongly suppressed. In particular, the electric-field enhancement factor is moderate and optical absorption is not very strong in this wavelength interval. For the opposite interval >600 nm, we observe a very different picture. Excitations in this spectral interval do not experience the inter-band broadening and the plasmonic effects become very strong. We see this behavior in Figure 3a and b. For the nanosphere (NST#9), the plasmon peak appears at ~ 520nm and the plasmon peak is relatively weak. For the NSTs, the plasmon resonance exhibits the typical red shift due to the geometrical factor. This red shift increases with the length of the NST tips. Simultaneously, the absorption cross section and the enhancement factor grow strongly with increasing the tip length. The rates of generation of hot electrons given by Equation (6) are directly proportional to the field-enhancement factor. Therefore, these rates should grow strongly as we move from the nanosphere to the NSTs with long tips. We see this behavior in Figure 3c. In Figure 3d and e, we show the other important properties of the NSTs. The surface-to-volume ratio is, of course, an increasing function of the spike length. The red shift of the plasmon peak in the NSTs comes together with the appearance of hot spots at the spikes of the NST since the surface charges in the dipolar plasmon become concentrated on the tips. Interestingly, our numerical calculations show that the plasmonic peaks of the NSTs in Figures 3b and c appear mostly owing to the tips (See Figure S3 in Supporting Information). Furthermore, Figure S3 shows the generation of hot electrons in the fractions of the tips in the Au NSTs. We see clearly that the related surface areas of the tips are relatively small (Table S1), but these areas with hot spots produce the majority of hot electrons in the whole NSTs. For example, one quarter of the tip in NST#4 produces approximately half of all hot carriers in this NC (Figure S3). This tells us that the hot spots at the tip regions play the major role in the process of generation of hot electrons in these nanostructures. In other words, the hot spots govern the plasmonic and electronic properties of the NSTs. The properties mentioned above will be important for following discussion.

We now compare the integrated properties of the NSTs and the spherical NP. For this, we will introduce a figure of merit for the efficiency of generation of hot electrons. First, it is convenient to integrate the generation rate with a waiting function. The natural waiting function is the solar spectrum. Second, we will divide the integrated rate by the NC volume. The NCs in our study play a role of catalyst and, therefore, the amount of catalysts used for generation of hot electrons plays a role.

We now apply Equation (6) to the case of the solar spectrum. The solar energy flux within the spectral interval $d\lambda$ is given by the standard equations:

$$dI = I_{solar} d\lambda,$$
$$I_{solar}(\lambda) = \frac{A}{\lambda^5} \frac{1}{e^{\frac{hc}{\lambda \cdot kT_{sun}}} - 1}, \quad T_{sun} = 5250K.$$

Then, the rate of generation of hot electrons by the photons within the spectral interval $d\lambda$ takes the form:

$$dRate_{high-energy} = W(\lambda) \cdot I_{solar} d\lambda,$$
$$W(\lambda) = \left( \frac{2}{\pi^2} \times \frac{e^2 E_F^2}{\hbar} \frac{(\hbar\omega - \Delta E_b)}{(\hbar\omega)^4} \int_{S_{NC}} \left| \frac{E_{normal}(\theta,\varphi)}{E_0} \right|^2 ds \right) \frac{2\pi}{c_0 \sqrt{\varepsilon_0}}.$$
(11)

Then, we integrate the rate (11) over the wavelength interval and normalize it by the NC volume. The resulting figure of merit takes the form:

$$Rate_{solar} = \frac{1}{V_{NC}} \int_{300nm}^{1239nm} d\lambda \times W(\lambda) \cdot I_{solar}(\lambda) \tag{12}$$

The units of the parameter $Rate_{solar}$ are $1/s \cdot nm^3$. Recently, we used the parameter $Rate_{solar}$ to describe experiments on photocatalytic activity of gold nanocrystals.[27] To understand the roles of different physical factors, we now write down an approximate expression of Equation (12),

$$Rate_{solar} \approx \frac{1}{V_{NC}} W(\lambda_p) \cdot I_{solar}(\lambda_p) \cdot \Delta\lambda_p = \frac{S_{NC}}{V_{NC}} Enh_S(\lambda_p) \cdot \lambda_p^4 \cdot (\hbar\omega_p - \Delta E_b) \cdot \Delta\lambda_p \cdot I_{solar}(\lambda_p) \tag{13}$$

where $\lambda_p$ and $\omega_p = 2\pi c_0/\lambda_p$ are the plasmonic wavelength and the plasmon frequency, respectively. The parameter $\Delta\lambda_p$ is the spectral width of the plasmon peak. To derive Equation (13), we assume that the plasmon peak gives the main contribution to the spectral integral (11). We see that the figure of merit $Rate_{solar}$ depends on several parameters including the surface-to-volume ratio, the field-enhancement factor at the plasmonic wavelength and the width of the plasmonic peak. In the following, we will use Equation (13) to interpret qualitatively our numerical results.

**Figure 4** shows the data for the solar rates $Rate_{solar}$ for various gold NSTs. For convenience, we normalize the data in Figure 4 to the rate of 15nm Au NP. The solar rate for such NP is

$$Rate_{solar,15nm-AuNP} = 7 \cdot 10^3 \frac{1}{s \cdot nm^3}$$

The above number was obtained for the total radiation intensity of

$$I_{tot} = \int_{300nm}^{1239nm} d\lambda \times I_{solar}(\lambda) = 10^2 \, W/cm^2$$

In Figure 4, we see clearly the trend of increase of the rate with the length of spikes. As we move from the spherical NP (#9) to the NST#1, we observe a 9-fold increase of the rate. The explanation is in the strong increase of the enhancement factor at the plasmonic wavelength (Figure 3b). The main physical reason, as we discussed above, is the red shift of the plasmon and the suppression of the interband transitions in gold. The surface-to-volume factor also plays some role, but it is not the main reason for the strong increase of the integrated solar rate of generation of hot electrons in the NSTs.

So far, we considered the NSTs with a fixed rounding radius of the tips. This radius was chosen to be 3nm. In Figure S4, we show the properties of NSTs with variable tip radii. As expected, we see that the position of the plasmon peak is very sensitive to this parameter. Simultaneously, the rate of generation of hot carriers does not vary much with the tip radius. Qualitatively, we can understand it in the following way: Of course, the sharper tips create stronger fields, but the volumes of such enhanced plasmonic fields are smaller in the sharper spikes. Therefore, the surface-integrated rates do not change strongly with the sharpness of the NST tip.

As we mentioned, the electromagnetic enhancement of the fields at the surface of a NC plays the leading role in the generation of hot electrons. This effect depends strongly on the width of the plasmon resonance $\Delta\lambda_p$. In general, the parameter, that determines the plasmon-peak width, is the Drude broadening $\gamma_D$.[45] This parameter enters the Drude term of the dielectric constant of a metal. One way to describe experimental data on plasmonic nanocrystals is to introduce an increased broadening into the empirical dielectric function:

$$\varepsilon_{metal, broadened}(\omega) = \varepsilon_{metal}(\omega) + \frac{\omega_p^2}{\omega(\omega+i\gamma_D)} - \frac{\omega_p^2}{\omega(\omega+i\gamma_{eff})}$$

where $\varepsilon_{metal}(\omega)$ is the experimental dielectric constant of a bulk metal. For the metals in this paper, we take the function $\varepsilon_{metal}(\omega)$ from Ref. [41]. For gold, the Drude parameters are: $\omega_p = 8.91\,eV$ and $\gamma_D = 0.078\,eV$. By changing the parameter $\gamma_{eff}$, we can control the broadening of the plasmon peak of a NC. In real NCs, experimental spectra typically show broader plasmon peaks as compared to calculations based on the bulk dielectric constant. For example, the experiment [46] was described with the parameter $\gamma_{eff} = 3\gamma_D$. **Figure 5** shows the data for the solar rates of the NSTs with different broadenings. As expected, the integrated rate decreases with increasing the plasmon broadening. This dependence is the following:

$$Rate_{solar} \propto \frac{1}{\gamma_{eff}} \qquad (14)$$

We can also see this behavior from Equation (13). The parameters in Equation (13) behave in the following way: $\Delta\lambda_p \propto \gamma_{eff}$ and $Enh_S(\lambda_p) \propto \gamma_{eff}^{-2}$. Then, using these dependences, we obtain Equation (14): $Rate_{solar} \propto Enh_S(\lambda_p) \cdot \Delta\lambda_p \propto \gamma_{eff}^{-1}$.

The case of gold NSTs shows the importance of plasmonic hot spots for generation of hot electrons. The NSTs exhibit almost isotropic optical response and, simultaneously, the dipolar plasmon peaks lies in the spectral interval > 600nm where the plasmon resonance does not suffer from the strong interband transitions. Therefore, the field enhancement factors and the rates in the NSTs should be strongly enhanced. We now compare the NSTs with the NRs (**Figure 6**). In contrast to the NSTs, the plasmonic NRs have strongly anisotropic optical responses. For convenience, we will denote the related rates as $Rate_{solar,x}$ and $Rate_{solar,y}$, for which the incident light should be polarized in the *x*- and *y*-directions. The averaged rate should be then calculated as

$$Rate_{solar} = \frac{Rate_{solar,x} + 2 \cdot Rate_{solar,y}}{3}$$

In Figure 6, the black curves show the data for the Au NRs for the two sets of geometries. We start with the transverse excitation geometry. If we vary the NR parameters keeping the volume constant (Figure 6a), the rate $Rate_{solar,y}$ stays nearly constant. For the geometries shown in Figure 6b, we see a decrease of the rate $Rate_{solar,y}$ with increasing the NR length for a constant NR width. The transverse plasmon resonance in the NRs is very weak as compared to the longitudinal plasmons in NRs. For the longitudinal rates of generation, we see strong enhancement with increasing the NR length. This is again because of the formation of strong plasmon peaks in the spectral region > 600 nm. In addition, the longitudinal plasmons in NRs create strong field enhancement in the whole volume of the nanocrystal. For efficient hot-electron generation, we need to have strong normal fields near the surface and, indeed, we observe such fields at the ends of the NRs (Figure 6, insets). Again, we see that the hot spot regions play the major role in the process of generation of hot electrons. The solar rates for the long Au NRs in Figure 6 show approximately 4-fold amplification in comparison with the spherical Au NPs. Interestingly, when we average the solar rate over the three directions of excitation, we do not see significant enhancement in the case of NRs, as compared with the spherical geometry (see solid curves in Figure 6). For the case (a) in Figure 6, we found 2-fold increase of the averaged rate for long NRs, whereas, for the case (b), the increase is not essential. It follows from the above results that the oriented NRs and polarized light are preferable to take advantage of the plasmonic hot spots in the NRs.

**4. Role of the Metal: Gold vs. Silver**

In this section, we will briefly discuss the role of the material system in generation of hot electrons. We see that silver NCs are much more efficient in generation of hot electrons (Figures 6 and S6). One of the reasons for this behavior is the plasmon broadening. Silver NCs typically exhibit narrower plasmon peaks compared to those of gold NCs. According to Equation (14), the integrated rate is inversely proportional to the Drude broadening parameter. From the fittings of the empirical dielectric constants [41], we obtain for gold and silver: $\gamma_{D,Au} = 0.078\ eV$ and $\gamma_{D,Ag} = 0.02\ eV$. We see that the Ag broadening parameter is much smaller than that for gold. Correspondingly, the field enhancement factors are much greater for the case of silver. In Supporting Information, we show the spectra for integrated surface fields of both Au and Ag NCs. Simultaneously, we see in Figure 6 that the integrated solar rates of the silver NCs are a few times larger. We, therefore, can conclude that silver is a better material for generation of hot electrons since silver NCs exhibit stronger plasmonic enhancements.

## 5. Efficiency of Hot Electron Generation and Quantum Parameter of Plasmons

In the final **Figure 7**, we show the quantum parameters for the process of optical absorption by different NCs, $Eff_{High-energy}$ and $QP$. These parameters are given by Equation (8) and (10). First, we see the size dependence of the efficiency: The efficiencies decrease with increasing the size of a NC. This is the typical behavior for the effect of generation of energetic electrons.[23,28] This effect is a surface effect coming from inelastic scattering of electrons by the surfaces. For large NCs, the quantum parameters of generation ($Eff_{High-energy}$ and $QP$) behave like $\propto S_{NC}/V_{NC} \propto 1/R_0$, where $R_0$ is the NC size. The material dependence of the quantum parameters in Figure 7 is very essential. Silver NCs have typically larger rates of generation compared to the case of gold. The physical reason is in the longer mean free path for electrons in Ag NCs and in the corresponding enhancement of quantum surface effects in silver. The mean free paths in Au and Ag estimated from the Drude scattering rates ($\gamma_{D,Au} = 0.078\ eV$ and $\gamma_{D,Ag} = 0.02\ eV$) are the following: $l_{mfp,Au} \sim 10\ nm$ and $l_{mfp,Ag} \sim 40\ nm$. We see that the mean free path in silver is much longer and, therefore, we expect stronger quantum effects in the silver NCs. In Figure 7, we also see that both parameters ($QP$ and $Eff_{High-energy}$) drop for short-wavelength photons due to the interband transitions in the noble metals, when the interband absorption intensity in such materials becomes very strong and dominates the picture. This happens for $\lambda < 500nm$ in the gold NCs and for $\lambda < 400nm$ in the case of silver. Overall, we see in Figure 7 that the NCs with sizes ~ 10-30 nm have large quantum contributions to the absorptions and demonstrate decent rates of generation of energetic plasmonic carriers. With increasing sizes of NCs, the quantum efficiency and the quantum parameter will decrease and approach zero in the bulk limit.

## 6. Conclusions

We have investigated the process of generation of hot plasmonic electrons in nanocrystals of various shapes. The nanostars show the best performance for this process. The nanostars have almost isotropic optical response and exhibit strong plasmonic hot spots, which serve as the sources of generation of energetic electrons. Since the hot-electron generation process comes from the amplified plasmonic fields, the nanocrystals with sharp plasmonic resonances show the best numbers for the integrated solar rates of generation of hot electrons. In that respect, the silver nanocrystals are more efficient than the gold ones. Another important factor is the interband transitions in a bulk metal; such transitions strongly reduce the plasmonic enhancement effect. To avoid the interband transitions, one should take nanocrystals with plasmons shifted to the red, such as nanostars or oriented nanorods. We also estimated the quantum efficiencies of generation of hot electrons for nanocrystals made of gold and silver. The quantum plasmonic effects and the process of hot-electron generation are intrinsically related. The generation of hot plasmonic electrons occurs via the quantum process of optical absorption at the surfaces of nanocrystals. Overall, we found that the quantum effects are very significant for the plasmonic responses of nanocrystals of typical sizes.

**Supporting Information**

Some details regarding the calculation of quantum absorptions and more data for optical properties of various types of nanocrystals. Supporting Information is available from the Wiley Online Library or from the author.

**Acknowledgements**

This work was supported by the Army Office of Research (MURI Grant W911NF-12-1-0407), by Volkswagen Foundation (Germany) and via Changjiang Chair Professorship (China). X.-T. K. was supported by the IFFS/UESTC oversea postdoc program and by Changjiang Scholar funding.

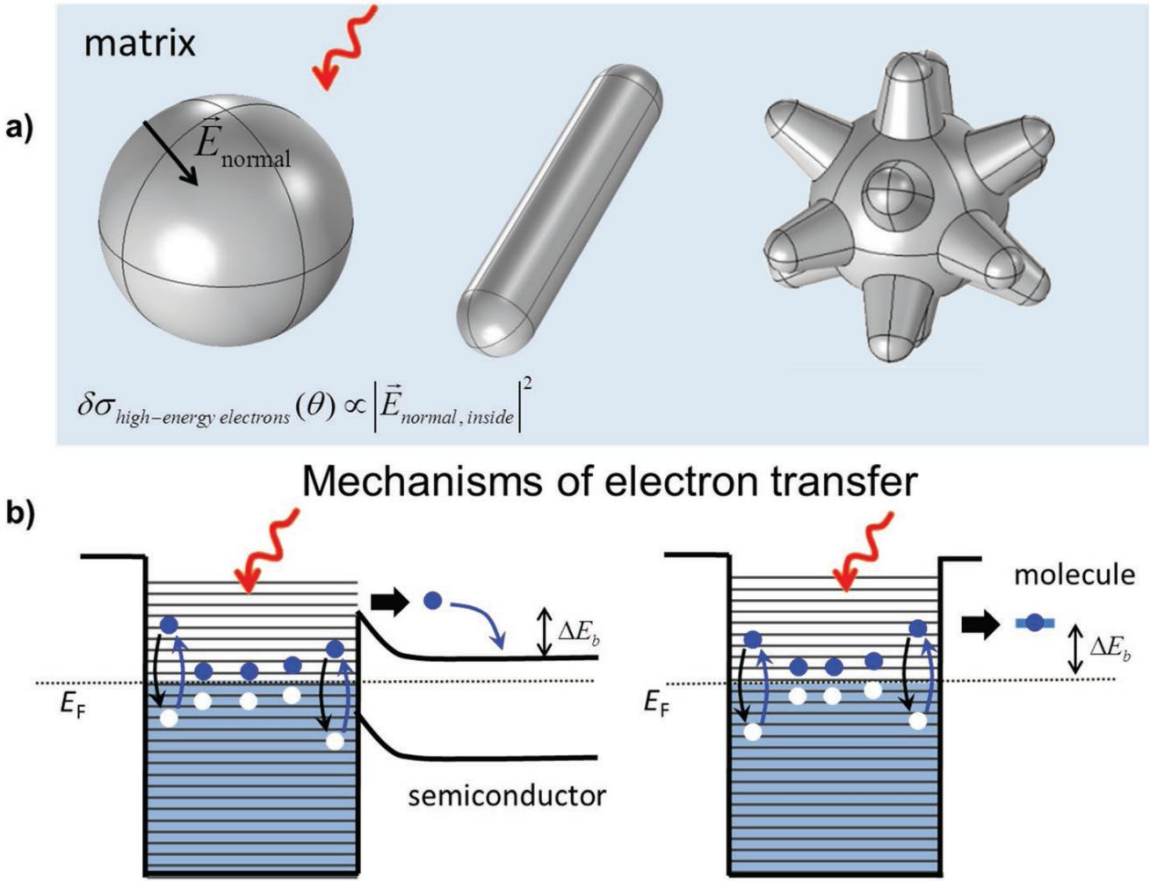

**Figure 1.** (a) Electromagnetic models of the nanocrystals considered in this study. The surface density of hot electrons is proportional to the normal component of the electric field near the surface. (b) Illustration of the excited states of plasmonic nanocrystals. Energetic electrons and holes are excited near the surfaces, whereas electrons and holes with low excitation energies are generated in the bulk. This figure also describes two mechanisms of extraction of hot plasmonic electrons from a nanocrystal: (1) Injection to a semiconductor and (2) Tunnel transfer to a molecule in a liquid.

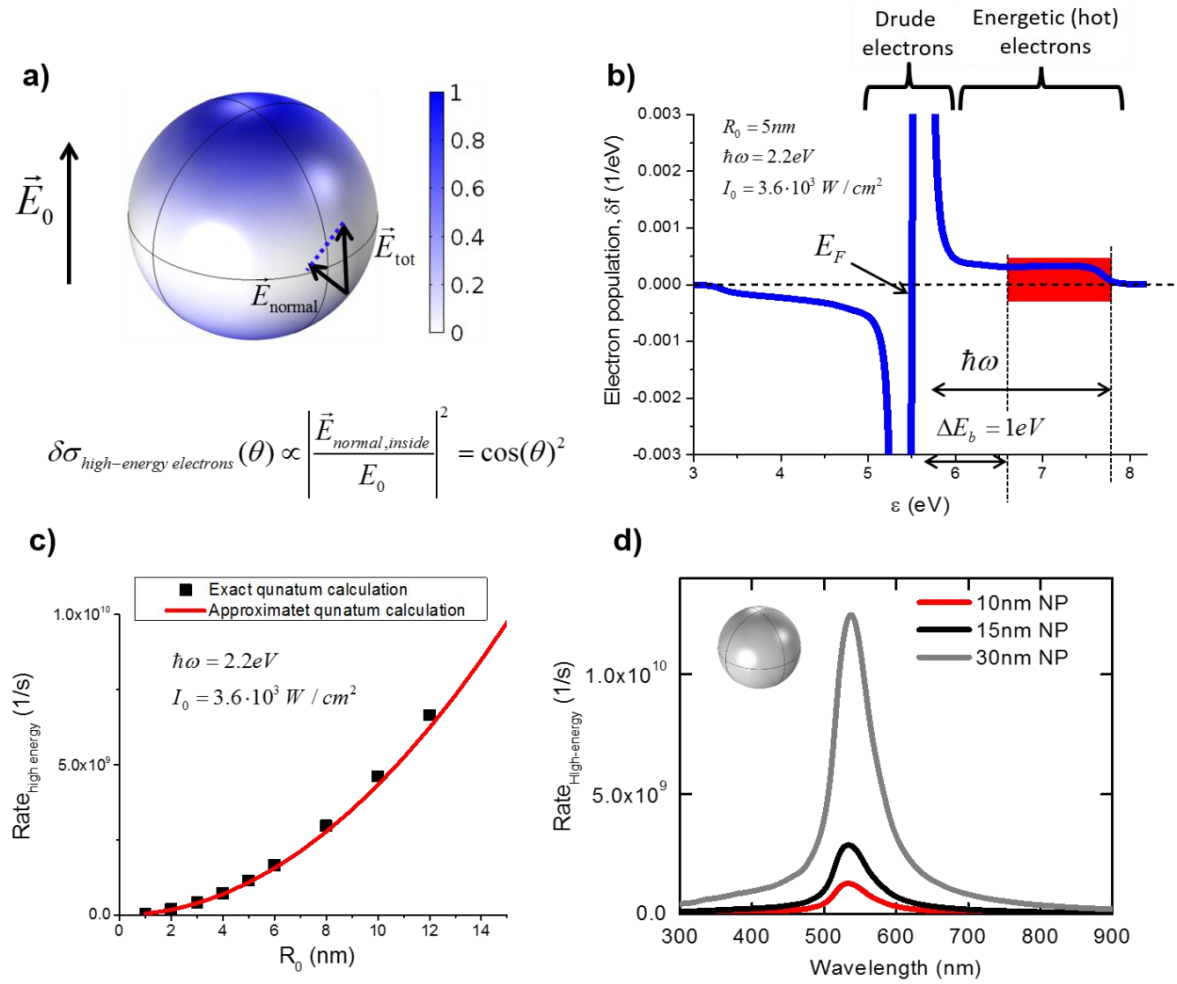

**Figure 2.** Properties of gold nanospheres. **(a)** Color map of the surface density of hot electrons generated by light in a nanosphere. **(b)** Nonequilibrium distribution of electrons in an optically-driven nanosphere. The red area indicates the interval of hot electrons with energies $\varepsilon > \Delta E_b$ that can be injected from the NC. **(c,d)** Rates of generation of hot electrons with $\varepsilon > \Delta E_b$ for nanospheres. The panel (c) shows the comparison between our simplified approach and the exact quantum calculation. The panel (d) shows the spectral properties of the hot-electron generation effect.

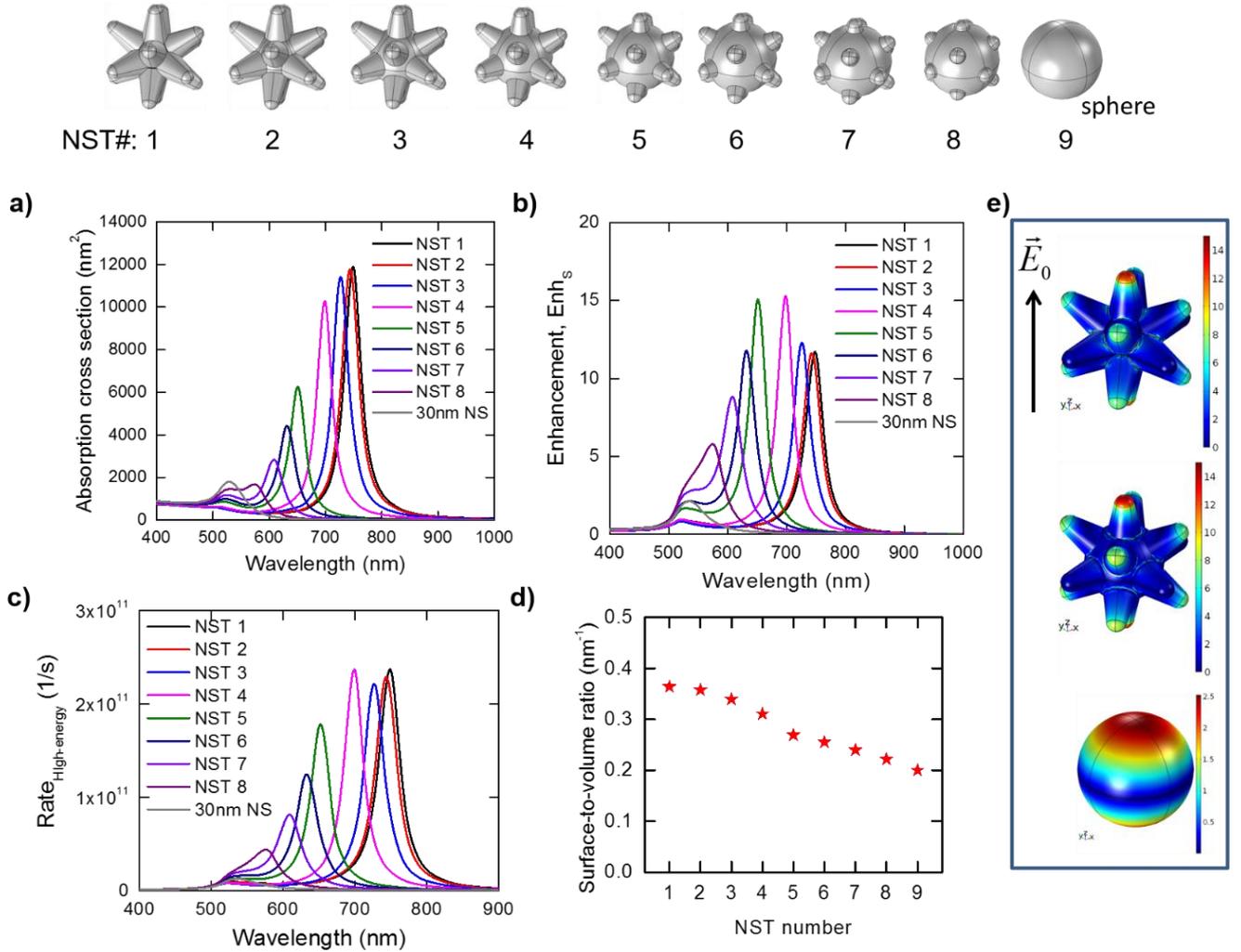

**Figure 3.** Optical and electronic properties of gold nanostars. **(a,b)** Absorption cross sections and surface-field enhancement factors. **(c)** Spectra for the rates of generation of hot electrons with large energies $E_F + \Delta E_b < \varepsilon < E_F + \hbar\omega$. **(d)** Surface-to-volume ratios for the NSTs. **(e)** Color maps of the normal electric fields near the surfaces for the two NSTs and for the sphere. The incident electric field is polarized in the *z*-direction. Upper insets: Models of the nanostars.

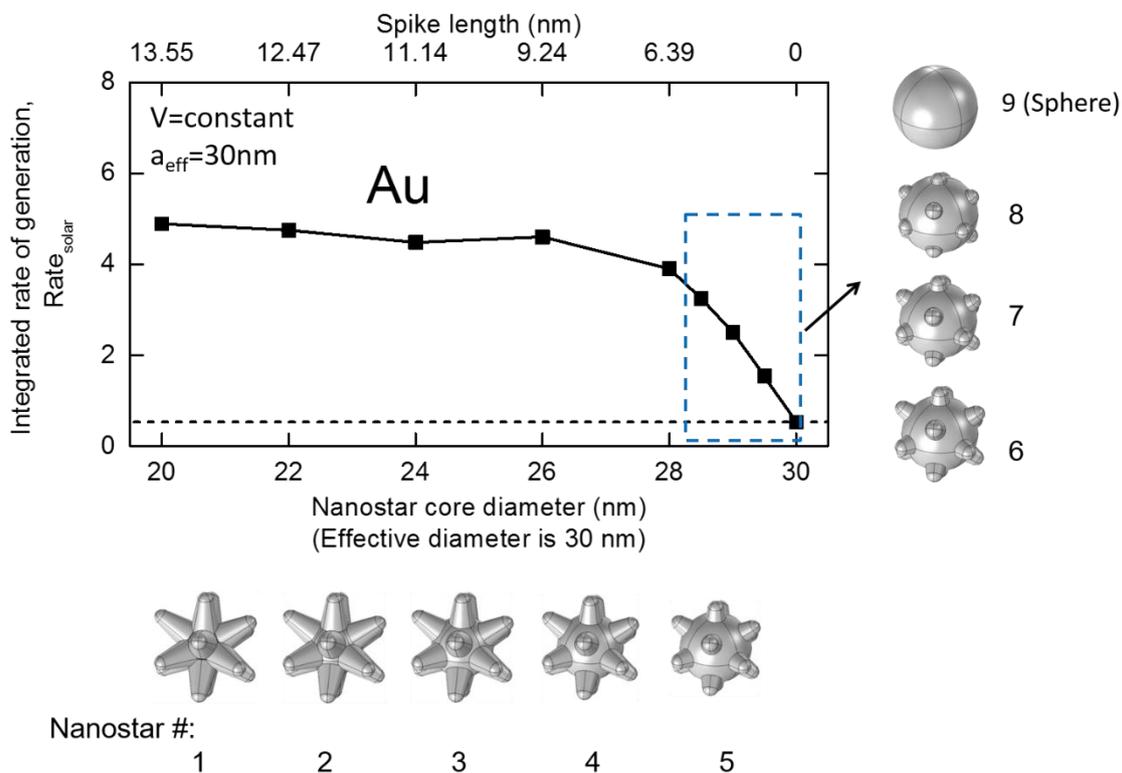

**Figure 4.** Normalized integrated rates of generation of hot electrons in the gold NSTs. The rates were weighted using the solar spectrum. The normalization was performed with respect to the integrated solar rate for the 15nm-AuNP. Insets: Models of the NCs.

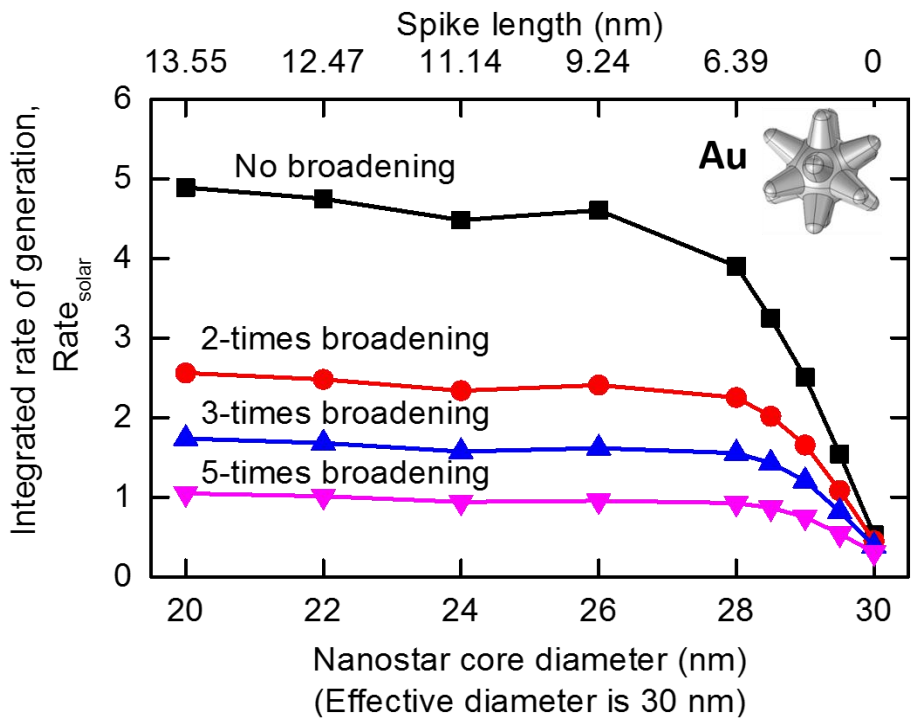

**Figure 5.** Normalized integrated rates of generation of hot electrons in the gold NSTs for different Drude broadenings. In this graph, we kept the volume of a NST constant and varied the spike length. The limiting case of the spherical geometry (30 nm core diameter) gives the smallest rate of generation.

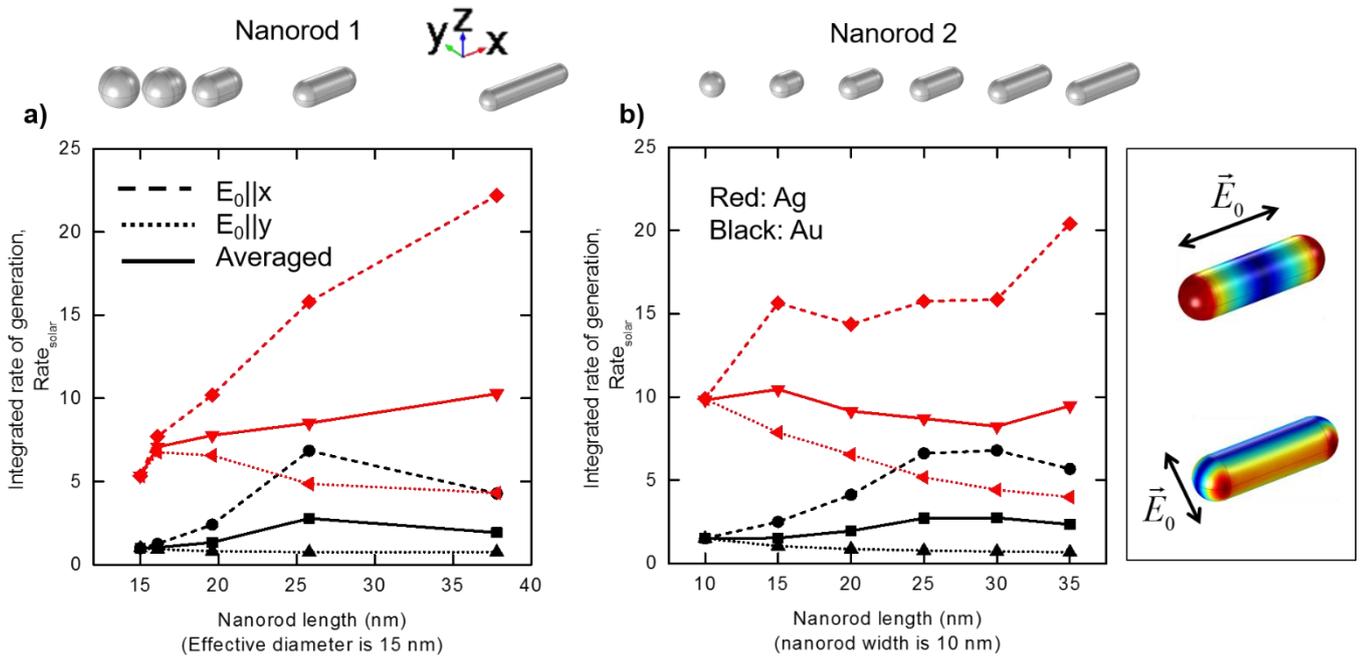

**Figure 6.** Normalized integrated rates of generation of hot electrons for the gold and silver nanorods. The figure shows the polarized and averaged rates. Upper insets: Models of the nanorods. Right insets: Color maps of the electric field normal to the surface inside the NR for the two polarizations of incident field.

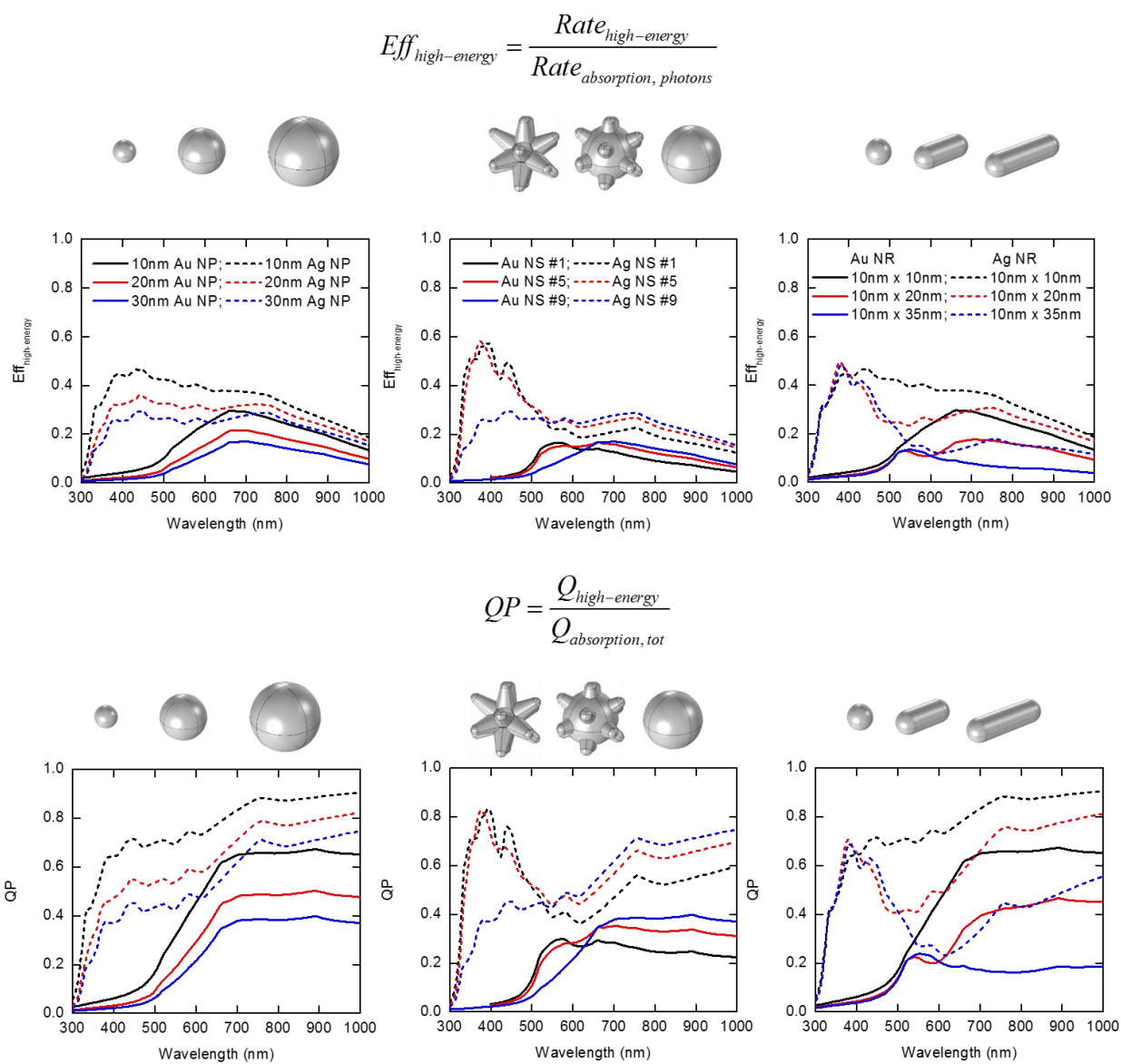

**Figure 7.** Efficiency of generation of hot electrons (upper panel) and the quantum parameter (lower panel). These graphs show the quantum parameters for the plasmonic NCs made of gold and silver and having various shapes.

# Supporting Information

**Plasmonic Nanostars with Hot Spots for Efficient Generation of Hot Electrons under Solar Illumination**

*Xiang-Tian Kong,\* Zhiming Wang, Alexander O. Govorov\**

## 1. Derivation of the Energy Dissipation Rate due to the Hot Electrons

In the kinetic approach, the dissipation of energy owing to relaxation of hot plasmonic electrons should be written as

$$Q_{tot} = \frac{1}{\tau_\varepsilon} \int_{|\varepsilon - E_F| > \delta E} d\varepsilon \cdot \varepsilon \cdot \delta f(\varepsilon)$$

where the integral is taken outside the energy region of the Drude electrons. The parameter $\delta E$ is given by the thermal energy $k_B T$. Then, we see that the function $\delta f(\varepsilon)$ is essentially flat in the interval of integration. Therefore, the integration gives

$$Q_{abs,quantum} = \frac{2}{\pi^2} \times \frac{e^2 E_F^2}{\hbar} \frac{(\hbar\omega)^2 - \delta E^2}{(\hbar\omega)^4} \int_{S_{NC}} |E_{normal}(\theta,\varphi)|^2 \times ds .$$

Since the optical energy significantly exceeds the thermal energy, i.e. $\hbar\omega \gg \delta E \sim kT$, we can simplify the above equation and arrive to the final equation

$$Q_{abs,quantum} \approx \frac{2}{\pi^2} \times \frac{e^2 E_F^2}{\hbar} \frac{1}{(\hbar\omega)^2} \int_{S_{NC}} |E_{normal}(\theta,\varphi)|^2 \times ds .$$

This equation is used in the main text.

## 2. Nanostar Model.

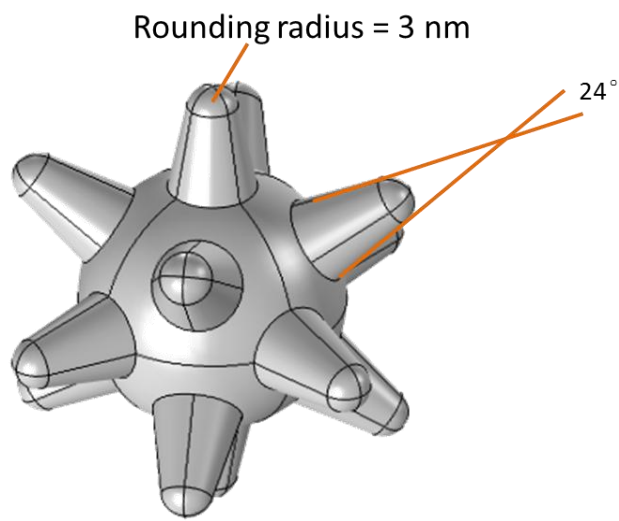

**Figure S1.** Geometry of a nanostar (NST #4) calculated in this study. The nanostar consists of 14 spikes. They are respectively in the directions of [±1, 0, 0], [0, ±1, 0], [0, 0, ±1], [1, 1, ± 1], [-1, 1, ± 1], [1, -1, ± 1], [-1, -1, ± 1]. The rounding radius of spikes for all the NSTs in the main text is 3 nm and the cone vertex angle is 24 degrees.

**Table S1.** Geometrical parameters of the nanostars (NSTs #1-#9) with the 3nm tip radius.

| Star No. | Core diameter (nm) | Spike length (nm) | Volume (nm³) | Surface area, spikes (nm²) | Surface area, core (nm²) | Surface area (nm²) |
|---|---|---|---|---|---|---|
| 1 | 20 | 13.5521 | 14137 | 5127 | 17 | 5144 |
| 2 | 22 | 12.4744 | 14137 | 4841 | 213 | 5054 |
| 3 | 24 | 11.1358 | 14137 | 4150 | 646 | 4796 |
| 4 | 26 | 9.2437 | 14137 | 3215 | 1176 | 4391 |
| 5 | 28 | 6.3964 | 14137 | 2027 | 1775 | 3802 |
| 6 | 28.5 | 5.4046 | 14137 | 1665 | 1942 | 3607 |
| 7 | 29 | 4.2464 | 14137 | 1272 | 2116 | 3388 |
| 8 | 29.5 | 2.8150 | 14137 | 829.8 | 2302 | 3132 |
| 9(sphere) | 30 | -- | 14137 | 0 | 2827 | 2827 |

## 3. COMSOL Models.

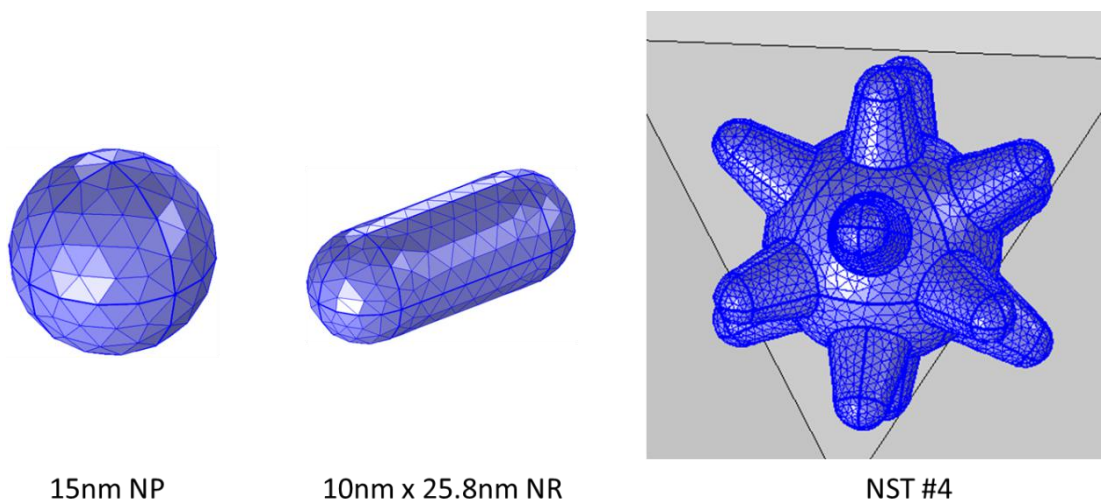

**Figure S2.** COMSOL models with meshes used in this study. The maximum mesh size for the metal region is 3 nm. The models were calculated using the RF module in COMSOL with solving the scattering fields. The electromagnetic boundary conditions at the metal-matrix interface in the COMSOL software are standard: Continuity of the normal components of the fields **D** and **B** at the metal surface and continuity of the tangential components of **E** and **H** at the metal boundary. Far-field scattering boundary conditions and perfect matched layers are used for the outer boundaries of the computational whole domains.

## 4. Role of the Spikes in Generation of Hot Electrons by the Au NSTs

$$Integral(\lambda) = \int_{S_{NC}} \left| \frac{E_{normal}(\theta,\varphi)}{E_0} \right|^2 ds$$

Solid: Total
Dashed: Contributed by the spikes
Dotted: Contributed by 1/2 spikes near the tips
Dot-dashed: Contributed by 1/4 spikes near the tips

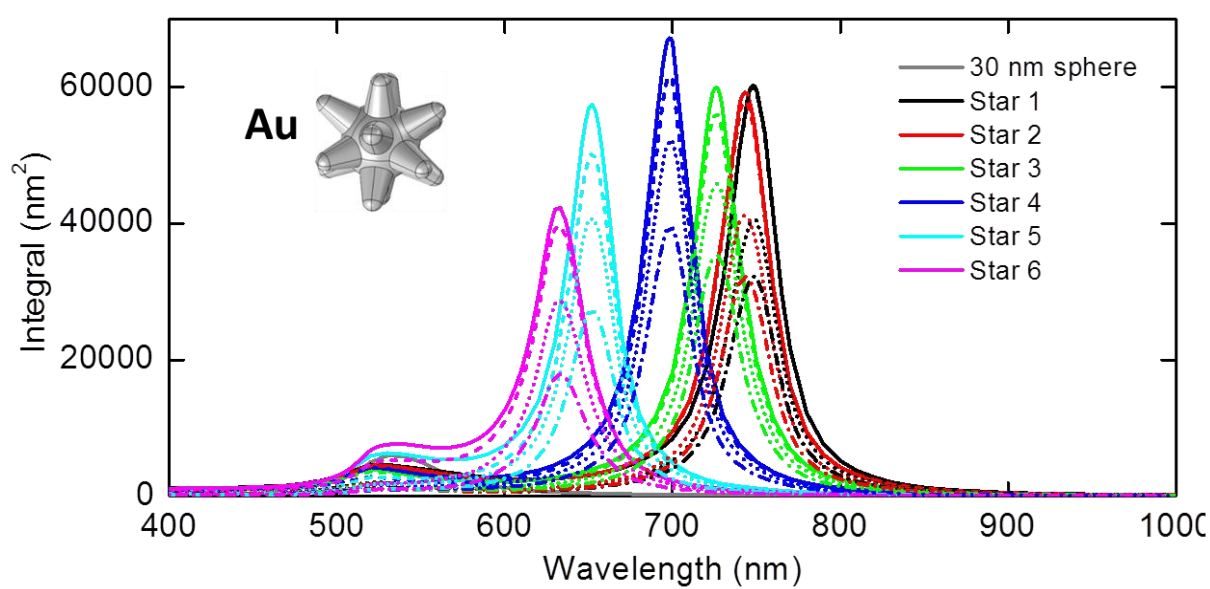

**Figure S3.** Enhancement factors of the surface electric fields of Au NSTs (#1 to #6) and 30nm Au NP. Solid lines: Contribution by all parts of the NSTs; Dashed lines: Contribution of the spikes; Dotted: Contribution of the halves of the spikes near the tips; Dot-dashed: Contribution of the quarters of the spikes near the tips.

# Au NSTs with differently rounded spikes

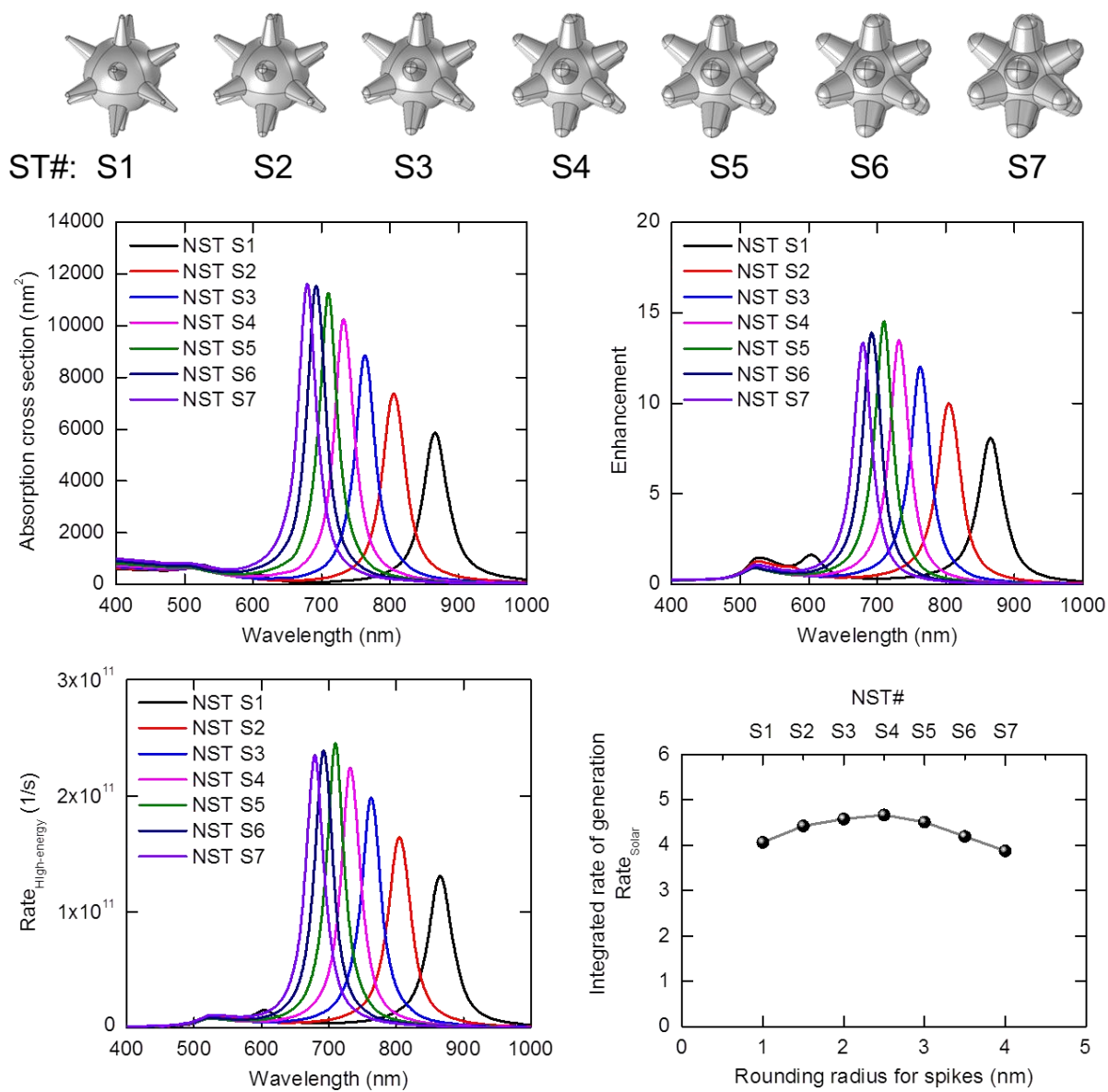

**Figure S4.** Optical properties of the Au NSTs with various tip radii. The spikes of the NSTs (#S1 to #S7) are rounded with radius from 1 nm to 4 nm. The detailed geometric parameters are given in Table S2.

**Table S2.** Geometrical parameters of the nanostars (#S1-#S7) with different tip radii.

| Star No. | Core diameter (nm) | Spike length (nm) | Rounding radius of spikes (nm) | Volume (nm$^3$) | Surface area, spikes (nm$^2$) | Surface area, core (nm$^2$) | Surface area (nm$^2$) |
|---|---|---|---|---|---|---|---|
| S1 | 26 | 10 | 1   | 10900 | 1781 | 1715 | 3496 |
| S2 | 26 | 10 | 1.5 | 11670 | 2198 | 1591 | 3789 |
| S3 | 26 | 10 | 2   | 12580 | 2632 | 1448 | 4080 |
| S4 | 26 | 10 | 2.5 | 13650 | 3085 | 1284 | 4369 |
| S5 | 26 | 10 | 3   | 14850 | 3560 | 1099 | 4659 |
| S6 | 26 | 10 | 3.5 | 16220 | 4059 | 891  | 4950 |
| S7 | 26 | 10 | 4   | 17730 | 4586 | 658  | 5244 |

## 5. NRs made of silver and gold

$$Integral(\lambda) = \int_{S_{NC}} \left| \frac{E_{normal}(\theta,\varphi)}{E_0} \right|^2 ds$$

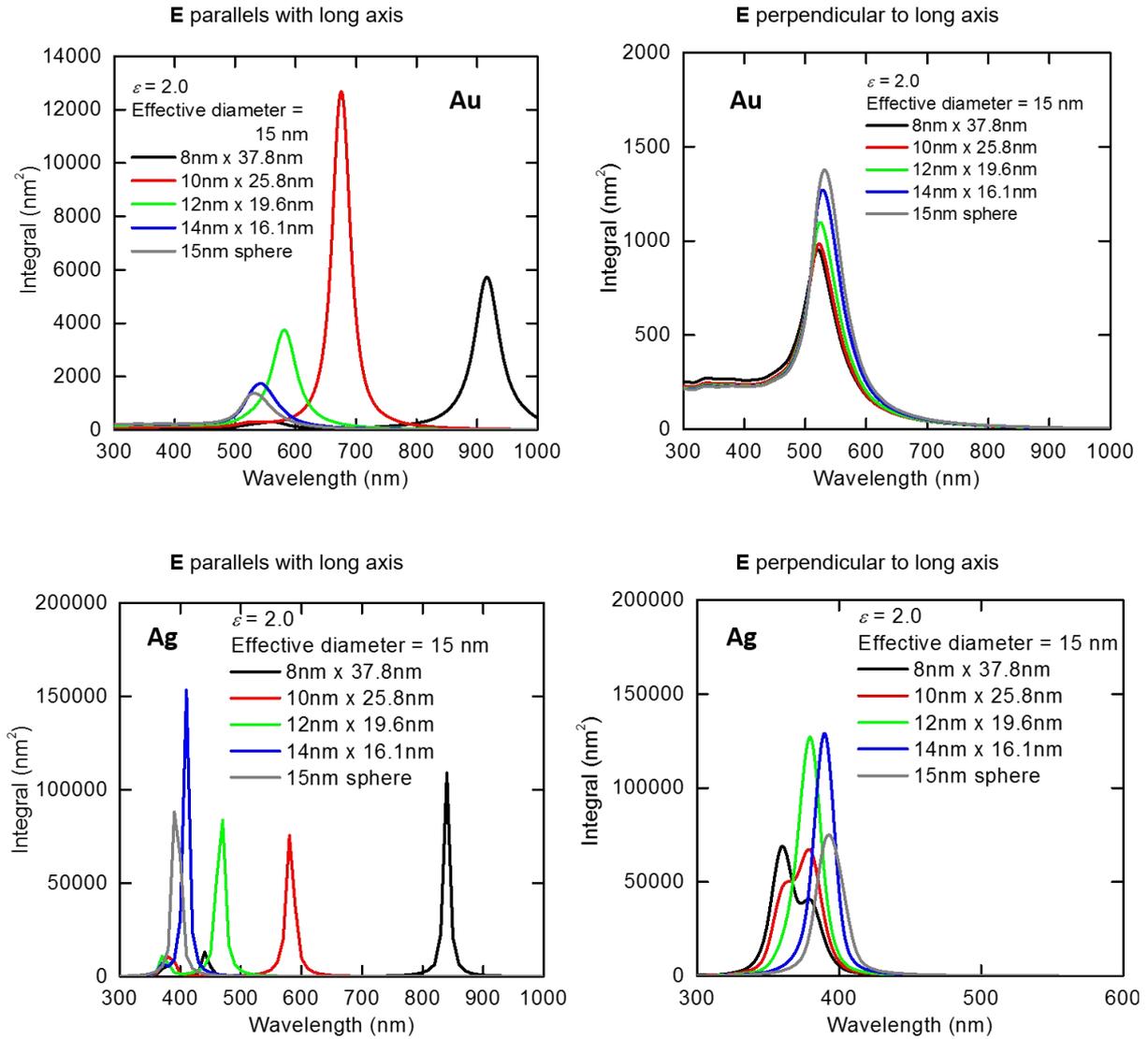

**Figure S5.** Enhancement factors for the surface electric fields of the Au and Ag NRs. The data are shown for the two polarizations of incident field. The effective diameter of the NRs is fixed at 15 nm. But, the aspect ratio of NRs is changed by varying the width of NRs from 8 nm to 15 nm.

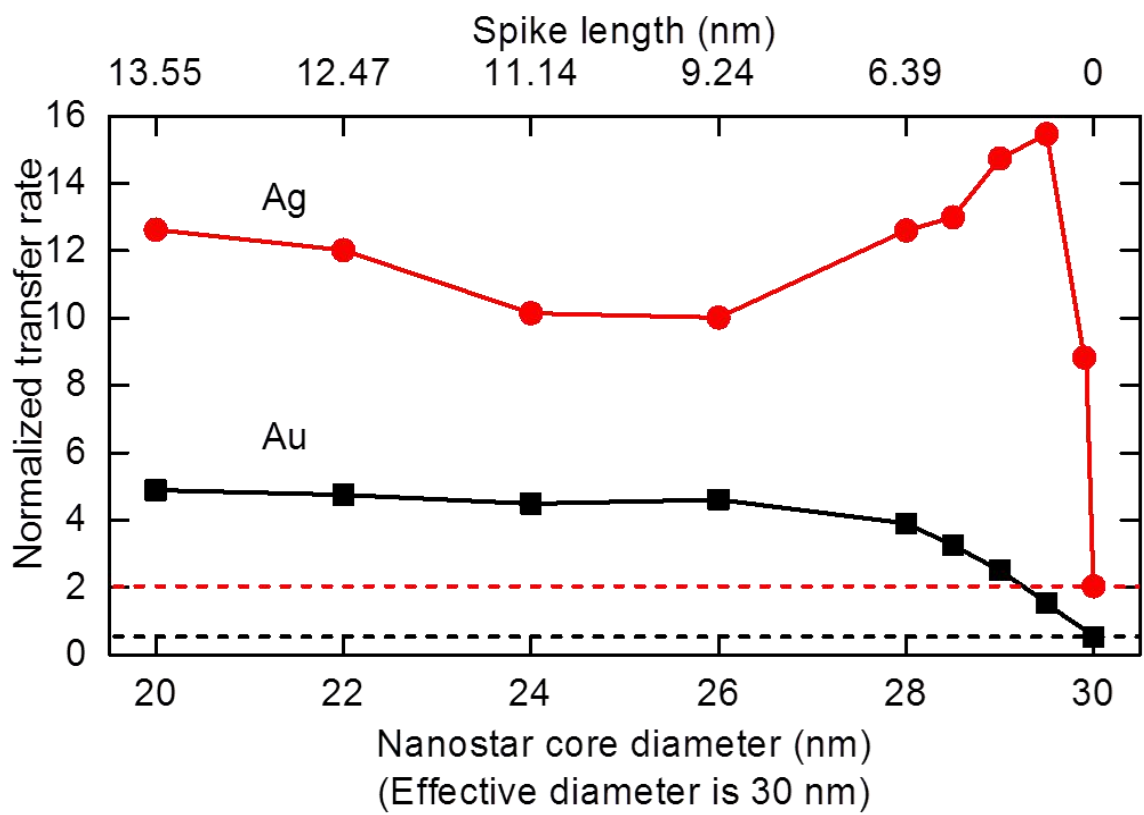

**Figure S6.** Comparison of the integrated solar rates for Au and Ag NSTs. The geometries of the NSTs are shown in Figure 4 in the main text. The horizontal dashed lines describe the spherical NCs. We again see that the silver NCs perform much better than the gold ones. The physical reasons for such behavior are addressed in the main text.